%% file: main.tex
\pdfoutput=1
\documentclass[format=acmsmall, review=false]{acmart}
\usepackage{booktabs}

\usepackage{microtype}
\usepackage[justification=centering]{caption}
\usepackage[flushleft]{threeparttable}
\usepackage{longtable}
\usepackage{multirow}
\usepackage{booktabs}
\usepackage{tabularx}
\usepackage{pifont}
\newcommand{\cmark}{\ding{51}}
\newcommand{\xmark}{\ding{55}}
\newcommand{\na}{\text{\sffamily n/a}}
\usepackage{tikz}
\newcommand{\pie}[1]{%
\begin{tikzpicture}
 \draw (0,0) circle (1ex);\fill (1ex,0) arc (0:#1:1ex) -- (0,0) -- cycle;
\end{tikzpicture}
}
\usepackage{etoolbox}
\usepackage{tikzpagenodes}
\def\ie{{i.e.},~}
\def\eg{{e.g.},~}

\usepackage{caption}
\usepackage{subcaption}
\usepackage[scaled=.75]{beramono}

\usepackage{listings}
\DeclareCaptionFont{black}{\color{black}}
 \iftrue
\newcommand{\gang}[1]{{\color{blue}{\bf GT comments:} #1}}
\else
\newcommand{\gang}[1]{}
\fi

\iftrue
\newcommand{\berkay}[1]{{\color{purple}{\bf BC:}\textbf{ #1}}}
\else
\newcommand{\berkay}[1]{}
\fi

\iftrue
\newcommand{\ef}[1]{{\color{red}{\bf EF:} #1}}
\else
\newcommand{\ef}[1]{}
\fi

\setcopyright{acmcopyright}
\setcopyright{acmlicensed}
\setcopyright{rightsretained}
\setcopyright{usgov}
\setcopyright{usgovmixed}
\setcopyright{cagov}
\setcopyright{cagovmixed}

\begin{document}
\title[\footnotesize{Program Analysis of Commodity IoT Applications for Security and Privacy: Challenges and Opportunities}]{Program Analysis of Commodity IoT Applications for Security and Privacy: Challenges and Opportunities}

\author{Z. Berkay Celik}
\orcid{1234-5678-9012-3456}
\affiliation{%
  \institution{Penn State University}
}
\email{zbc102@cse.psu.edu}

\author{Earlence Fernandes}
\affiliation{%
  \institution{University of Washington}
}
\email{earlence@cs.washington.edu}

\author{Eric Pauley}
\affiliation{%
  \institution{Penn State University}
}
\email{eap5377@psu.edu}

\author{Gang Tan}
\affiliation{%
  \institution{Penn State University}
}
\email{gtan@cse.psu.edu}

\author{Patrick McDaniel}
\affiliation{%
  \institution{Penn State University}
}
\email{mcdaniel@cse.psu.edu}

\renewcommand{\shortauthors}{B. Celik et al.}

\input{abstract.tex}

%
%


%
%
\settopmatter{printacmref=false}
\setcopyright{none}

\makeatletter
\renewcommand\@formatdoi[1]{\ignorespaces}
\makeatother
\renewcommand\footnotetextcopyrightpermission[1]{} 

\maketitle
\fancyfoot{}
\thispagestyle{empty}

\input{introduction.tex}
\input{background.tex}
\input{IoTProgrammingPlatforms.tex}
\input{ProgAnalysis.tex}
\input{StudyOfIoTSystems.tex}
\input{conclusions.tex}

\section{Acknowledgements}
The authors thank Xiaolei Wang, Dongrui Zeng and Leonardo Babun for helpful discussions about this work. Research was supported in part by the Army Research Laboratory, under Cooperative Agreement Number W911NF-13-2-0045 (ARL Cyber Security CRA) and the National Science Foundation Grant No. CNS-1564105. The views and conclusions contained in this document are those of the authors and should not be interpreted as representing the official policies, either expressed or implied, of the Army Research Laboratory or the U.S. Government. The U.S. Government is authorized to reproduce and distribute reprints for Government purposes notwithstanding any copyright notation here on. Earlence Fernandes is supported by the University of Washington Tech Policy Lab and the MacArthur Foundation.

\bibliographystyle{ACM-Reference-Format}
\bibliography{iot-survey.bib}

\end{document}

%% file: abstract.tex
\begin{abstract}
Recent advances in Internet of Things (IoT) have enabled myriad domains such as smart homes, personal monitoring devices, and enhanced manufacturing. 
IoT is now pervasive---new applications are being used in nearly every conceivable environment, which leads to the adoption of device-based interaction and automation. 
However, IoT has also raised issues about the security and privacy of these digitally augmented spaces.
Program analysis is crucial in identifying those issues, yet the application and scope of program analysis in IoT remains largely unexplored by the technical community. 
In this paper, we study privacy and security issues in IoT that require program-analysis techniques with an emphasis on identified attacks against these systems and defenses implemented so far.
Based on a study of five IoT programming platforms, we identify the key insights that result from research efforts in both the program analysis and security communities and relate the efficacy of program-analysis techniques to security and privacy issues.
We conclude by studying recent IoT analysis systems and exploring their implementations. Through these explorations, we highlight key challenges and opportunities in calibrating for the environments in which IoT systems will be used. 
\end{abstract}

%% file: introduction.tex
\section{introduction}
The introduction of IoT devices into public and private spaces has changed the way we live. For example, home applications that integrate smart locks, thermostats, switches, surveillance systems, and appliances allow users to monitor and interact with their living spaces from anywhere. While industry and users alike have embraced IoT, concerns have been raised about the security and privacy of digitally augmented spaces~\cite{ronen2017iot, fernandes2016security, ho2016smart}. IoT environments necessarily have access to functions that, if abused, would put user security at risk, \eg unlock doors when the user is not at home or create unsafe conditions by turning off the heat in cold weather~\cite{soteria}. In addition, these networked systems have access to private data that, if leaked, would cause privacy issues, \eg information about when the user sleeps or who and when others are at home~\cite{saint-taint-analysis}.

Driven by consumer concerns, one of the central criticisms of IoT is that existing platforms lack the essential tools and services to analyze security and privacy. Such criticisms have not gone unnoticed. Recent technical community efforts have proposed a range of tools to identify sensitive data leaks in IoT apps~\cite{saint-taint-analysis,fernandes2016flowfence}, while others have focused on improving IoT safety and security~\cite{soteria, jia2017contexiot, tian2017smartauth, wang2018fear}. Works in this area use program-analysis techniques to design and build algorithms that identify vulnerabilities and dangerous behavior within a targeted IoT programming platform. These works motivate our work to study security and privacy issues in IoT that are solved by program-analysis techniques.

While thematically similar to program analysis in mobile apps and other domains, from our study of five major IoT programming platforms (Samsung's SmartThings, Apple's HomeKit, OpenHAB, Amazon AWS IoT, and Android Things), we have found that IoT programming platforms present unique characteristics and challenges in program analysis when compared to other platforms~\cite{saint-taint-analysis}. First, in the case of Android, a well-defined intermediate representation (IR) is available, and analysis can directly analyze IR code. However, IoT programming platforms are diverse, and each uses its own programming language. Second, IoT integrates physical processes with digital connectivity through a diverse set of devices, each of which has a different set of internal device states (\eg door locked/unlocked); thus identifying security and privacy issues through these physical states is quite subtle. For example, an adversary can break into a home by changing the thermostat temperature value that causes the windows to open once the temperature reaches a threshold value. Lastly, each IoT programming platform has its own idiosyncrasies that can pose challenges to program analysis. For instance, the SmartThings platform allows apps to perform call by reflection and make web-service requests; each of these features makes program analysis more difficult and requires special treatment. Due to these domain-specific challenges, ensuring the safety, security, and privacy of IoT systems is not a trivial endeavor.

In this work, we present security and privacy issues in IoT that motivate program analysis techniques. We contrast program analysis in IoT with other domains, demonstrating key differences that complicate analyses. We first study five IoT programming platforms to gain insights into the structure of their apps. We then present IoT-specific issues that require program-analysis techniques within an IoT app or multiple-apps colocated in an environment. We focus on areas that prior research has addressed and others that remain open problems. Lastly, we demonstrate a number of IoT program idiosyncrasies that require special treatment and present several general precision requirements for IoT code analysis, by providing examples from IoT apps. We conclude by studying a representative set of recent IoT analysis systems from literature. Our study serves as a guideline for researchers and provides insights into the design and implementation of IoT program analysis for security and privacy. In this work, we explore the following:

\begin{itemize}  \setlength\itemsep{1em}
\item  We conduct a study of five major IoT programming platforms to understand their program structures. We map their program structures to a sensor-computation-actuator idiom that includes the common building blocks of IoT apps.

\item We present IoT-specific issues, IoT program idiosyncrasies and general precision requirements for IoT app analysis with examples from 230 SmartThings IoT apps. We discuss the problems that have already started to study by the research community and the areas that need attention. In highlighting open issues, we draw insights and motivate future work.

\item We study six IoT analysis systems from literature for security and privacy that incorporate program-analysis techniques. We measure their ability to analyze IoT apps and evaluate their approaches to IoT-specific issues. We note that we limit our analysis to publications that use program analysis for IoT security and privacy and were published at a major venue.
\end{itemize}

\vspace{3pt}\noindent\textbf{Scope.} This work is at the intersection of three domains: IoT programming platforms, program analysis, and security and privacy. IoT programming platforms provide a software stack to develop applications that monitor and control devices. Program analysis includes the techniques used for analyzing the behavior of an IoT app or multiple-apps in an environment. Security and privacy cover the objective of the program-analysis techniques to identify potential security and privacy issues. We begin below by giving an overview of IoT systems and program structures of IoT platforms.

%% file: background.tex
\section{Background} 
We start with an overview of how IoT systems structure their design (Section~\ref{sec:iot-architectures}). We then present recent research on IoT security and privacy (Section~\ref{sec:iot-securityandprivacy}). As IoT is a diverse domain, we focus on consumer IoT, which has the largest number of applications and the most significant market~\cite{iotappsplatforms}. 

\begin{figure}[t!]
\begin{center}
\includegraphics[width=0.7\columnwidth]{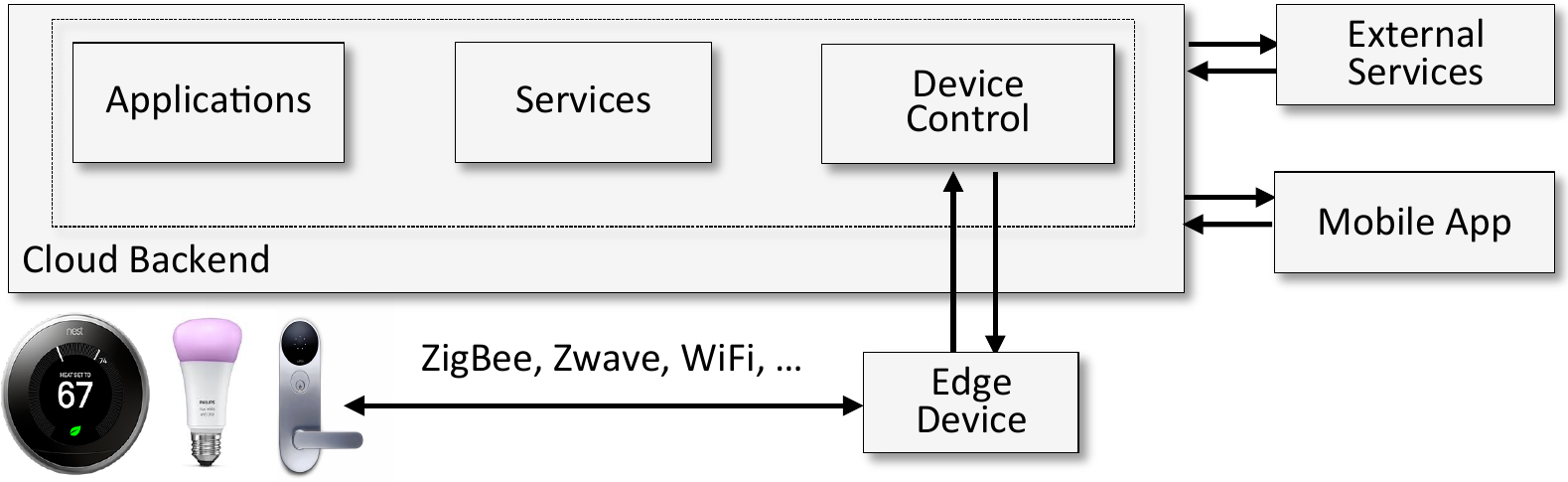}
\caption{An example architecture of edge-based IoT system.}
\label{fig:IoTArchitecture}
\end{center}
\end{figure}

\subsection{An Overview of IoT System Architectures}
\label{sec:iot-architectures}
IoT systems integrate physical processes with digital connectivity. These systems are used to achieve simple tasks such as motion-activated light switches, as well as complex tasks such as controlling the traffic lights in a smart city. Regardless of their purpose and complexity, IoT systems often structure their architecture from bottom to top with (1) devices, (2) connectivity protocols, and (3) IoT programming platforms (See Figure~\ref{fig:IoTArchitecture}). These systems often use an edge device as a centralized gateway that connects devices in a physical environment, use a cloud back-end to synchronize device states and provide interfaces for remote control and monitoring of devices. 

Devices are equipped with embedded sensors and actuators that interact with a physical environment. Sensors collect physical states and send events to other devices, the hub, or the cloud. These events are processed and used to actuate the devices. For example, a presence sensor detects a presence event and communicates with a switch (actuator) that turns on the lights. We note that a mobile phone or even a coffee machine can be a sensor as long as it can gather information about its environment. Protocols are used to establish communication between heterogeneous devices and network endpoints. These protocols are selected according to the requirements of the environment such as low power or non-lossy connection. For instance, the Bluetooth Low Energy (BLE) protocol is used for short-range communication and is extremely energy efficient.

IoT programming platforms deliver app-specific services by managing devices and their interactions. They also enable crucial functions such as data collection, control, and interoperability. In recent years, several IoT programming platforms are emerged in a wide range of domains:  Apple's HomeKit~\cite{apple}, OpenHAB~\cite{openhab}, Samsung's SmartThings\cite{samsung} for smart home, Android Sensor API~\cite{androidapi}, Google Fit for wearables~\cite{googlefit}, ThingWorx~\cite{thingsworx} for aerospace, Eclipse Kura~\cite{eclipsekura} for general-purpose solutions, and FarmBeats~\cite{vasisht2017farmbeats} for agriculture. These platforms offer web-based environments and tools that enable developers to write applications used to create custom automations. Applications use a diverse set of languages and execute in a variety of environments (\eg the cloud or a local hub). Further, in some IoT platforms, applications are written in a Domain Specific Language (DSL)~\cite{openhab} and applications run in a sandbox for performance and security purposes~\cite{smartThings-documentation}.

\subsection{IoT Security and Privacy}
\label{sec:iot-securityandprivacy}
The growth of IoT devices has had profound impacts in settings such as automotive industry~\cite{kirk2015cars}, aviation~\cite{comitz2016aviation}, smart homes~\cite{fernandes2016security}, medical wearables~\cite{yang2014health}, agriculture~\cite{vasisht2017farmbeats}, and smart cities~\cite{zanella2014internet}. While industry and users have widely embraced IoT, concerns are also raised about the security and privacy of these digitally augmented spaces~\cite{saint-taint-analysis, laz16, orc16}, and led to fervent calls for restrictions and rigorous standards regulating their uses~\cite{eca17}. Indeed, the risk of IoT failures and misuse has proven to be real. Vulnerable and faulty devices can lead to everything from privacy violations (\eg compromised baby-monitors~\cite{wax14}) to vehicle crashes and monetary theft~\cite{vee16}. In other domains, failures can lead to serious health consequences (\eg failed IoT pacemakers~\cite{tay16}) or even result in catastrophic environmental disasters (\eg pipeline explosions~\cite{jab15}). 

In response to the security and privacy threats in IoT, most attempts to date aim to improve perimeter defenses that harden the IoT infrastructure against attacks using firewalls~\cite{kubler2015standardized}, intrusion detection systems~\cite{zarpelao2017survey}, access control policies~\cite{WeijiaIoTaccess}, and software patches~\cite{leiba2018incentivized}. Other efforts have explored vulnerability analysis within specific IoT devices and IoT programming platforms. Oluwafemi et al.~\cite{oluwafemi2013experimental} investigated the security risks in smart lights controlled by compromised automation systems, and Ho et al.~\cite{ho2016smart} studied the vulnerabilities of smart locks. Fernandes et al. discovered design flaws in permission control of the SmartThings IoT platform~\cite{fernandes2016security}, and Xu et al.~\cite{xu2014security} surveyed the security problems in IoT hardware design. These works have found that applications can be easily exploited to gain unauthorized access to control devices and leak sensitive information of users and devices. Past analysis of IoT devices and environments have also focused on securing an IoT app through source code analysis. Most previous studies rely on techniques designed for mobile phone security~\cite{EnckTaintDroid, ZhuTaint, GuTaint, clause2007dytan, ArztFlowdroid, reaves2016droid}. For instance, some systems infer an app's context to enforce permissions based on that context through runtime prompts~\cite{jia2017contexiot} or asking users for authorization through an interface~\cite{tian2017smartauth}.

There are also several recent surveys on IoT security and privacy, which differ in scope and focus from this work. These surveys centered on the security of emerging IoT devices and protocols. Alwari et al. proposed a methodology to analyze security properties for home-based IoT devices~\cite{sokSP}. Roman et al. performed a study on reported IoT attacks and defenses~\cite{roman2013features}. Others focused on security analysis of IoT architectures~\cite{zhang2017understanding}, available security solutions~\cite{jing2014security}, and privacy threats~\cite{ziegeldorf2014privacy, acar2018peek}. 
This work studies the space of IoT application security and privacy research through program-analysis techniques. Those seeking a survey of IoT more broadly can look to many recent papers covering this rapidly-developing area~\cite{xu2014security, oluwafemi2013experimental, ho2016smart, sivaraman2015network, yu2015handling, fernandes2017internet}. 

%% file: IoTProgrammingPlatforms.tex
\input{tablePlatformStudy.tex}

\section{IoT Programming Platforms}
IoT platforms provide a software stack used to develop apps that monitor and control IoT devices. In 2018, there are more than hundreds of IoT platforms in the marketplace~\cite{iotappsplatforms}. We focus on five IoT platforms that have the largest market share, Samsung's SmartThings, OpenHAB, Apple's HomeKit, Android Things, and Amazon AWS IoT. We present a survey of these IoT platforms to gain insights into the structure of their apps (Section~\ref{sec:platformOverview}).  Table~\ref{table:app-properties} summarizes our study. Our survey was performed by reviewing the platforms' official documentation, running their example IoT apps, and analyzing their app construction logic. A broad investigation showed that IoT platforms use similar programming structures and the differences lie only in the communication protocols between IoT devices and edge systems. Therefore, we generalize their programming structures to the sensor-computation-actuator idiom, which is used to model an IoT app (Section~\ref{sec:iotAppStructures}).

\subsection{Overview of IoT Programming Platforms}
\label{sec:platformOverview}

\noindent\textbf{Samsung's SmartThings} consists of a hub, apps, and the cloud back-end~\cite{smartThings-review, soteria}. The hub controls the communication between connected devices, cloud back-end, and mobile apps. Apps are developed in the Groovy language (a dynamic, object-oriented language) and executed in a Kohsuke sandboxed environment. The cloud backend creates SmartDevices that act as software proxies for physical devices and also runs the apps. The permission system in SmartThings allows a developer to specify devices and user inputs required for an app at install time. Devices in SmartThings have capabilities (\ie permissions) that are composed of \emph{actions} and \emph{events}. Actions represent how to control or actuate device states and events are triggered when device states change. SmartThings apps control one or more devices (See Listing~\ref{listing:smartThings}). Apps subscribe to device events or other pre-defined events such as the icon-clicking event, and an event handler is invoked to handle it, which may lead to further events and actions.

\begin{lstlisting}[float=t!, language=Java, caption = SmartThings IoT application structure, label=listing:smartThings]
/* Metadata describing how app is shown in UI */
definition(...)
/* Run-time binding of devices and user inputs */
preferences {...}
/* Predefined methods for updating, initialization, and installation of an app */
def updated() {...}
def initialize() {...}
def installed() {
    subscribe(device, "device event", handler)
}
def handler() {
// Computation and actuators.
}
\end{lstlisting}

\vspace{3pt}\noindent\textbf{OpenHAB} is an open-source automation platform built in the Eclipse IDE~\cite{openhab}. It provides vendor- and technology-agnostic support for various devices specifically designed for home automation. OpenHAB provides flexible device integration and rules to build automated tasks. Similar to the SmartThings platform,  the rules are implemented through triggers to react to the changes in the environment (See Listing~\ref{listing:openHAB}). For instance, event-based triggers listen to events generated from devices; timing-based triggers respond to special times (\eg midnight); system-based triggers run with certain system events such as system start and shutdown. The rules are written in a Domain Specific Language (DSL) based on the Xbase language, which is similar to the Xtend language~\cite{efftinge2012xbase}. Users can install OpenHAB apps by placing them in the rules folder of their installation directories or by downloading from the Eclipse IoT Marketplace~\cite{openHABThirdParty}.

\begin{lstlisting}[float=t!, language=Java, caption = OpenHAB IoT rule structure, label=listing:openHAB]
rule "<RULE_NAME>"
when
    /* Define events */
    <TRIGGER_CONDITION>
    [or <TRIGGER_CONDITION2> [or ...]]
then
    /* Computation and actuators */
    <SCRIPT_BLOCK>
end
\end{lstlisting}

\vspace{3pt}\noindent\textbf{Apple's HomeKit} is a development kit that manages and controls compatible smart devices~\cite{apple}. The HMHomeManager class describes a set of homes (locations).  An HNHome class defines each house and each room within that set. Each room may include a different number of accessories (HMAccessory). Accessories represent the physical devices. Each accessory supports a service (HMService), similar to the device capabilities in SmartThings, such as unlocking the door. Services of an accessory are organized as HMServiceGroup which defines accessory services as an individual asset. Accessories are also formed based on the zones (HMZone). This enables developers to group home locations such as the basement, living room and kitchen.  Lastly, each service includes specific characteristics (HMCharacteristic), which describes the services such as a Boolean (locked or unlocked) or floats (the thermostat temperature value). Developers write scripts to specify a set of actions, triggers, and optional conditions to control HomeKit-compatible devices. HomeKit applications can either be written in Swift or Objective C. Users can install HomeKit apps using the Home mobile application provided by Apple~\cite{appleMarket}.

\begin{lstlisting}[float=t!, language=C++, caption = Apple HomeKit IoT application structure, label=listing:appleHomeKiT]
/* Create a home with properties such as the rooms */
private func initialHomeSetup() {...}
/* UI setup for devices and user inputs via HMAccessory */
override func tableView(...) {...}
/* Computation and actuators */
func eventsActions() {
/* Create an HMCharacteristicEvent that invokes when an event happens */

/* Use HMEventTrigger to create predicates that must be met before an action is executed */ 

/* Use executeActionSet to execute all the actions in a specified action set (actionSets) */
}
\end{lstlisting}

\begin{lstlisting}[float=t!, language=SQL, caption = AWS IoT rule structure, label=listing:amazonAWS]
  "sql": "SELECT events from devices WHERE conditions",
    "description": " Rule description",
    "actions": 
    [
      {
        /* Take actions when an incoming message meets the conditions defined in the rule. */ 
      }
    ]
}
\end{lstlisting}

\vspace{3pt}\noindent\textbf{Amazon Web Services (AWS) IoT}  provides communication between smart devices and the AWS Cloud~\cite{amazonIoT}. Connected devices transmit their states to AWS IoT Core. However, optional IoT hubs can be installed to help bridge the connection or add additional use cases. For instance, a home user can use Amazon's Alexa voice assistant to control smart devices. A device shadow service abstracts the physical device and saves the state of the devices for use by other devices or services. Applications are deployed to AWS IoT Core as companion apps and server apps. Companion apps connect to devices through the cloud. For example, a mobile app might use AWS IoT to unlock a smart lock at the user's request. Server apps monitor and control many connected devices. For instance, a fleet operation app might use AWS IoT to map thousands of vehicle locations in real-time. AWS IoT implements interfaces to create and interact with the devices. For instance, the AWS IoT API offers a set of interfaces to develop apps using HTTP requests, and the AWS SDK wraps the HTTP APIs and enables developing apps using language-specific APIs in languages such as Java and C. Furthermore, AWS IoT supports SQL-like rules, which are used for filtering messages sent to AWS IoT Core and transfer them to other devices or a AWS cloud service (See Listing~\ref{listing:amazonAWS}). A rule can use data from many devices and perform a set of actions at the same time.

\vspace{3pt}\noindent\textbf{Android Things} is an Android-based embedded operating system that enables developers to build smart devices and IoT apps~\cite{androidThings}. It is built on the core Android app programming stack: official software development kit, Android Studio, and Google Play services. Android Things uses the same lower layers of the stack as Android. For the app framework, the Things Support Library incorporated while specific Android APIs are omitted in Android Things. This library integrates with new hardware types that are not found on conventional Android devices. An app running on an embedded device creates an activity as the main method in its manifest file when the device boots (See Listing~\ref{listing:androidThings}). The apps then monitor device state changes through listeners. When a device event happens, a callback is triggered to implement  app functionality.

\begin{lstlisting}[float=t!, language=Java, caption =  Android Things IoT application structure, label=listing:androidThings]
public class ClassName extends Activity{
    protected void onCreate(...) {
      // Detect events and register a callback to take actions when the event happens 
      registerGpioCallback(GpioCallback callback) {...}
    }
    /* Close connections and nullify hardware references */
    protected void onDestroy(...) {...}
    /* Callback method invoked from onCreate() */
    private callback(...) {
      // Computation and actuators
    }
}
\end{lstlisting}

\subsection{Generalizing IoT Application Structure}
\label{sec:iotAppStructures}
\begin{figure}[t!]
\begin{center}
\includegraphics[width=0.68\columnwidth]{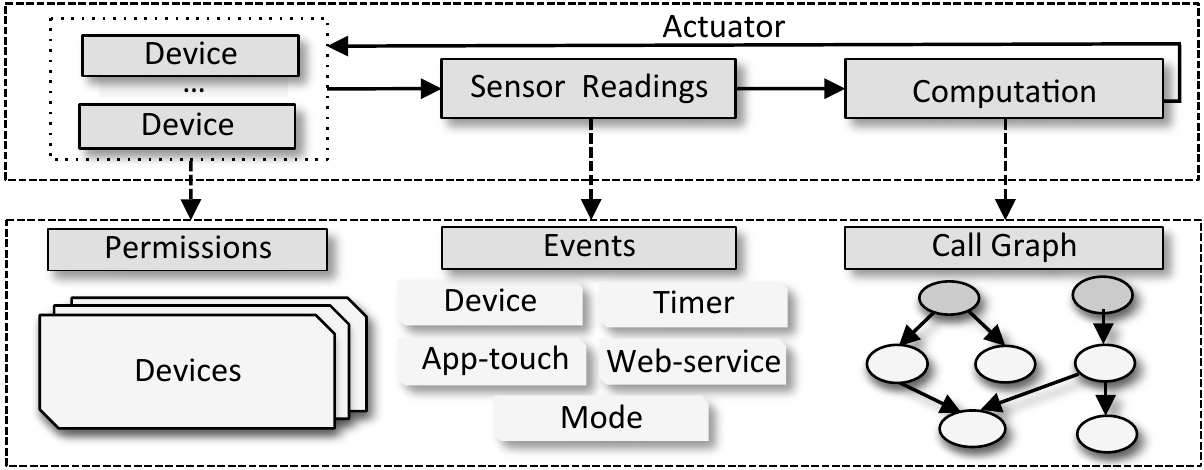}
\caption{Mapping IoT application structures to the sensor-computation-actuator idiom.}
\label{fig:program-structure}
\end{center}
\end{figure}

A broad investigation of dominant IoT platforms shows that IoT systems structure their apps' design around the \emph{sensor-computation-actuator} idiom regardless of their purpose and complexity~\cite{saint-taint-analysis, soteria}. Therefore, the source code of an IoT app can be translated to a platform-agnostic structure with three types of common building blocks as shown in Figure~\ref{fig:program-structure}: (1) \emph{Permissions} grant access to devices  and user inputs used in the app to implement the app functionality; (2) \emph{Events} reflect the association between sensor readings and actuators: when a sensor reading is triggered, a device is actuated; and (3) \emph{Call graphs} represent the relationship between entry points and call-sites in the app.

Permissions are granted when an app is installed or updated. This is where various types of devices and user inputs are described and granted access. Apps can only interact with devices for which they have been given permission. Devices have capabilities of \emph{actuators} and \emph{sensor readings}. Actuators represent the actions that a device can do and sensor readings represent the state information of devices. Actuators and sensor readings are not one to one. While a device may support many sensor readings, it may have a limited number of actuators, \eg a door may have opening, opened, closing and closed sensor readings, but has only open and close actuators. 

Events connect particular sensor readings and handler methods. That is, when an event through a sensor reading is triggered by a device, an associated event handler of an app is invoked. Event handlers may actuate changes in the state of the devices. For instance, when a motion sensor reports a motion-active event, an app may invoke an event handler to actuate a light switch from off to on. We found that events are not limited to device events; while different IoT platforms name these differently, we call them \emph{abstract events} and classify them into four different groups~\cite{saint-taint-analysis}\footnote{During the time we wrote the paper, some platforms started supporting additional abstract events. One such example is OpenHAB's system events, which are triggered when a system boots up or shuts down. We refer readers to platform documentation for a complete set of events a platform supports.}: (1) \emph{Timer events}; event-handlers are scheduled to take actions within a particular time or at pre-defined times (\eg an event-handler is invoked to take actions after a given number of minutes has elapsed or at specific times such as sunset); (2) \emph{App touch events}; for example, some action can be performed when the user taps on a button in an app; (3) \emph{External events}; IoT programming platforms may allow an app to be accessible over the web; this enables external entities (\eg If This Then That (IFTTT)~\cite{ifttt}) to make requests to the app and get information about or control end devices; (4) what actions get generated may also depend on \emph{mode events}, which are behavior filters that are used to automate device actions; for instance, an app running in ``home'' mode turns off the alarm and turns on the alarm when it is in the ``away'' mode. 

An IoT app does not have an entry method (\ie main method) due to its event-driven program structure. Apps implicitly define entry points by subscribing events through event handler methods. An app may have multiple entry points by subscribing to multiple events. Additionally, apps often call other functions in event handlers to implement logic, send messages, or log device events to a database. A call graph is used to represent this control-flow relationship between a particular event handler and other functions the event handler invokes.

%% file: tablePlatformStudy.tex
\begin{table*}[t!]
\caption{Summary of studied IoT programming platforms (as of July 2018).}
\label{table:app-properties}
\def\arraystretch{1.3}
\setlength{\tabcolsep}{2pt}
\resizebox{\textwidth}{!}{%
\begin{threeparttable}[b]
\begin{tabular}{|l|c|c|c|c|c|c|c|} 
\hline
\multicolumn{1}{|c|}{\textbf{IoT Platform}} & \textbf{Architecture\textdaggerdbl} & \textbf{App execution} & \textbf{Abstract events} & \textbf{Sandboxing*} & \textbf{Official apps} & \textbf{3rd-party apps}  & \textbf{Programming lang.} \\ \hline \hline
SmartThings& Hub & Hub/Cloud & \cmark & \cmark & \cmark~\cite{Official} & \cmark~\cite{Community}  & Groovy\\ \hline
OpenHAB&Hub& Hub & \cmark & \pie{180}\tnote{$\bullet$} & \cmark~\cite{openHABOfficial} & \cmark~\cite{openHABThirdParty}  & Xtend-based DSL\\ \hline
Apple's HomeKit & Hub & Hub &  \cmark & \cmark & ${\na}^+$& \na & Swift/Objective C\\ \hline
Android Things & Cloud & Cloud &  \cmark & \cmark &\cmark~\cite{androidThingsOfficial}& \na & Java\\ \hline
Amazon AWS IoT & Both & Cloud &  \cmark & \cmark & \na& \na & SQL-like, (Java, Python, C)\textdagger    \\ \hline
\end{tabular}
\begin{tablenotes}[para, small]
    \item [\textdaggerdbl] means whether devices connect to hub or cloud.
    \item [*] means sandboxing is enforced or not. 
    \item [$\bullet$] \pie{180} means it is optional.
    \item [\textdagger] means that programming language depends on SDKs. 
    \item [$+$] \na~ means that there is no official app repository managed by the IoT platform.
\end{tablenotes}
\end{threeparttable}
}
\end{table*}

%% file: ProgAnalysis.tex
\section{Program Analysis of {IoT} Apps}
\label{sec:prog-analysis}
Program-analysis techniques operate on IoT app source code to achieve a variety of goals, such as understanding apps' security. In this section, we begin by identifying common program analysis goals (Section~\ref{sec:analysisPurpose}), followed by challenges in IoT program analysis (Section~\ref{sec:IoTchallenges}). In the next section, we classify the program analysis issues into three groups and discuss each of them (Section~\ref{sec:prog-analysis-criteria}). In the following, we explore contemporary approaches to understand security and privacy threats.

\subsection{Goals of Analyses}
\label{sec:analysisPurpose}
We first discuss several common goals of performing program analysis on IoT apps. Many of these goals remain open problems; thus understanding the goals can guide future work.

\vspace{3pt}\noindent\textbf{Sensitive Data Leaks.}  IoT devices have access to data that can be intensely private \eg the door is locked or unlocked and users are present home or away~\cite{saint-taint-analysis}. IoT platforms ensure only coarse-grained access controls to sensitive information and provide limited controls over how that information is used. For instance, if a user lets an app access the energy meter, the user cannot know if the app will send the energy usage to the app developer, advertisers, or to any other entity.  

\vspace{3pt}\noindent\textbf{Abuse Prevention.} IoT apps necessarily have access to functions that if abused would put the user's safety and security at risk, \eg unlock doors when the user is not at home~\cite{ho2016smart} or create unsafe or damaging conditions by turning on a smart oven~\cite{denning2013computer}. Therefore, it is crucial to prevent IoT apps from abusing device capabilities by ensuring those apps operate devices according to a set of security, safety, and functional properties~\cite{soteria}.

\vspace{3pt}\noindent\textbf{Permission Misuse.} The permission model of an IoT platform defines an app's access to sensitive actions such as device state changes. However, IoT apps may misuse permission models. This can happen for two main reasons. First, a permission model may be coarse-grained and conflate permissions of devices; for example, an app granting the permission to a door lock grants access to both door lock, and door unlock actions, even though the app may only need the privilege of locking the door~\cite{lee2017fact}. Second, an app may trick users to acquire unneeded and dangerous device permissions; for example, a smoke-alarm app may request the permission of a security camera to disable it, even though the app does not need the permission to function~\cite{tian2017smartauth}.

\vspace{3pt}\noindent\textbf{Data Provenance.} As IoT apps perform increasingly diverse activities, attacks and misconfigurations require investigation. To address this, provenance systems use program instrumentation that aims to collect IoT app information to construct complete and accurate app behavior. After that, they aggregate that information into a data structure such as provenance graphs for forensics and system diagnosis. For instance, a provenance system designed for IoT apps may provide complete history of device actions and events, which can be used to identify the cause of an attack~\cite{wang2018fear, IoTDots}.

\begin{figure}[t!]
\begin{center}
\includegraphics[width=0.78\columnwidth]{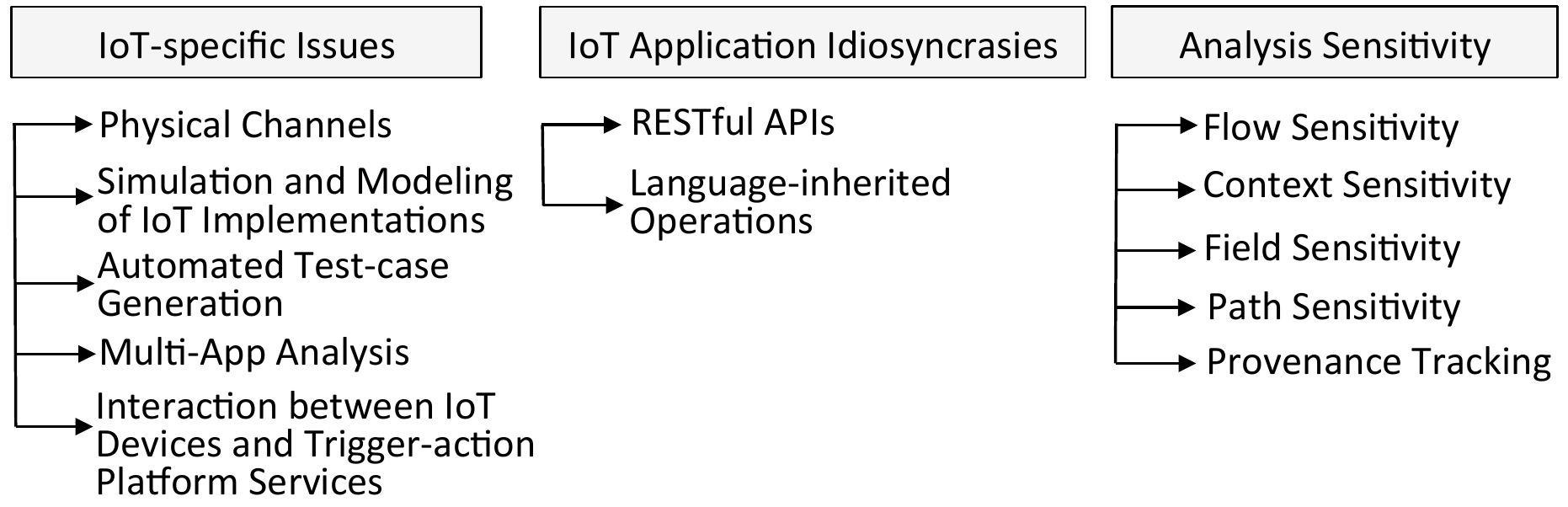}
\caption{Categorization of issues in IoT program analysis  discussed in Section~\ref{sec:prog-analysis-criteria}.}
\label{fig:classification}
\end{center}
\end{figure}

\subsection{Issues in IoT Program Analysis} 
\label{sec:IoTchallenges}
Program analysis has been applied, either statically or dynamically, to many different settings such as mobile apps. From our study of five IoT platforms, we found that IoT platforms possess a few unique characteristics and issues when compared to other platforms. 

First, in the case of Android, a well-defined Intermediate representation (IR) is available, and analysis can directly analyze IR code. For instance, popular analysis frameworks including Soot~\cite{soot} and WALA~\cite{wala2016watson} that have been used to analyze Android app source code provide libraries to convert Dalvik bytecode to the Jimple IR~\cite{bartel2012dexpler}, to construct call graphs~\cite{lhotak2003scaling}, and to perform inter-procedural dataflow analysis via graph reachability~\cite{bodden2012inter}. However, IoT programming platforms are diverse, and each uses its own programming language. Therefore, the analysis must capture the event-driven nature of IoT apps, and perform analysis on it.

Second, IoT apps control physical hardware peripherals and drivers. Consequently, IoT apps have qualitatively different vulnerabilities resulting from handling physical processes such as temperature, smoke, motion, humidity, water leak, and luminance. For instance, an adversary might misuse the capability of an IoT device through physical channels to achieve a damaging effect. To illustrate, we consider an app that grants permissions to a smart light, which supports color and intensity capabilities. The app may strobe light at a frequency and change colors to various shades, which could trigger seizures in users who have photosensitive epilepsy~\cite{ronen2016extended}. 

Third, IoT apps may interact with each other when they are co-located in an environment. The interaction between apps, among others, may happen when a device action executed in an app's event handler is used as an event to trigger another app's event handler~\cite{soteria}. For instance, two apps interact with each other when the  ``switch off'' action of an app is used as a ``switch turned-off'' event in another app. The interactions among apps may lead to undesirable device states causing security and safety violations and exposing users to risks such as a locked door when there is a fire.

Fourth, trigger-action platforms such as IFTTT~\cite{ifttt}, Zapier~\cite{zapier}, and Microsoft Flow~\cite{MicrosoftFlow} are increasingly used to bridge the divide between physical (\eg IoT devices) and digital (\eg e-mail services, social media platforms) processes. These platforms allow users to use rules that connect the events and actions of IoT devices with the events and actions of digital services. For example,  a user may use a rule that posts a Tweet when she turns on the light in the living room, and similarly, another rule logs the user's presence to a spreadsheet file when the front door is unlocked. This inter-tangled environment expands the interactions among devices to online services~\cite{celikIoTGuard, surbatovich2017some}; for example, an IoT app that subscribes to the switch ``turn-on'' event interacts with a trigger-action platform rule that ``turns on'' the switch when the user is tagged in a photo on Facebook.

Lastly, each IoT platform has its own idiosyncrasies that can pose challenges to program analysis~\cite{saint-taint-analysis}. For instance, SmartThings IoT apps written in the Groovy programming language that allows apps to perform call by reflection and allows web-service apps; each of these features makes analysis more difficult and requires special treatment.

\begin{figure}[t!]
\begin{center}
\includegraphics[width=0.78\columnwidth]{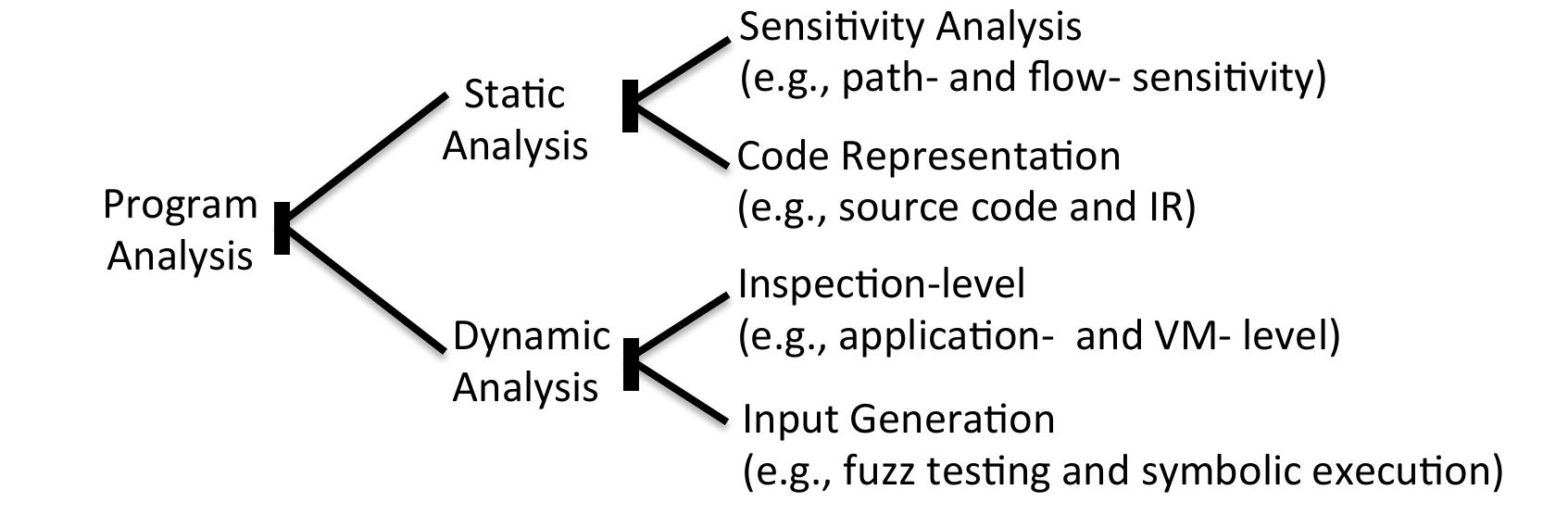}
\caption{Categorization of issues based on program analysis type.}
\label{fig:StaticDynamic}
\end{center}
\end{figure}

\section{Analysis of IoT Programs}
\label{sec:prog-analysis-criteria}
Based on our discussion in the previous Section, we split the analysis characteristics and challenges of IoT apps into three groups as shown in Figure~\ref{fig:classification}. In this section, we begin by presenting IoT-specific analysis issues (Section~\ref{sec:iot-specific-issues}). We then detail IoT application idiosyncrasies (Section~\ref{sec:IoT-app-idiosyncrasies}). Lastly, we present general precision requirements for IoT code analysis (Section~\ref{sec:analysis-sensitivity}). 

\vspace{3pt}\noindent\textbf{Type of Program Analysis.} Previously covered issues can be addressed through static or dynamic code analysis, and in some cases, issues are related to both. For example, path sensitivity is not an issue in dynamic analyses since they follow execution paths; they instead suffer from coverage problems. We split these issues based on the analysis type as shown in 
Figure~\ref{fig:StaticDynamic}. In static analysis, the source code of an app is analyzed without running it, and in dynamic analysis, the code is run, possibly under-instrumented conditions, to see if there are likely problems~\cite{aho1986compilers}.  Static analysis benefits from analyzing the complete source code whereas, in dynamic analysis, only a portion of the code is executed; thus analysis results are limited to observed executions. Furthermore, static analysis may lead to over-approximations by generalizing all possible behaviors of a program, risking false positives~\cite{ernst2003static}. For instance, an analysis tool detects a sensitive data leak through a piece of code in an IoT app which is not executable at run-time. A dynamic analysis may under-approximate because the execution inputs of a program are often incomplete; thus the analysis may produce false negatives. For example, an analysis tool may miss vulnerabilities or malicious behaviors in an IoT app at run-time.

\vspace{3pt}\noindent\textbf{Example Code Blocks.} During our discussion, we will provide example code blocks obtained from our analysis of 230 SmartThings apps~\cite{saint-taint-analysis}. We primarily reference SmartThings because a large number of open-source market apps are available, and it has a detailed, publicly available documentation that helps validate our findings~\cite{smartThings-documentation}. In late 2017, we obtained 168 official (vetted) apps from the SmartThings GitHub repository~\cite{Official} and 62 community-contributed third-party (non-vetted) apps from the SmartThings community forum~\cite{Community} (See Table~\ref{table:sample-apps}).~\footnote{The apps are available at our IoTBench test-suite repository~\cite{iotbench}.}  These apps were selected to include various IoT devices and contexts that encompass diverse real-life use cases.

\input{tableExampleApps.tex}

\subsection{IoT-specific Analysis Issues}
\label{sec:iot-specific-issues}
IoT apps possess unique characteristics and challenges in terms of program analysis when compared to other platform apps. In this section, we enumerate five challenges that are mainly due to the capabilities provided by IoT platforms to the apps. 

\vspace{3pt}\noindent\textbf{Physical Channels.} IoT devices integrate physical processes into digital connectivity. Misuse of physical processes allows an app to deviate from a device's intended functionality to achieve an unexpected effect. We give three examples of physical processes that lead to security and privacy issues: (1) data-leaks through side-channels, (2) health-related risk through device functionality misuse, and (3) safety issues through indirect physical interactions.

We demonstrate the first two examples with an app that grants access to a light device. The light has the capability to change color, hue, saturation, and intensity level. The first example is an app that creates a side-channel by changing the light intensity to notify an adversary or another app when the households are sleeping or not at home ~\cite{saint-taint-analysis, jia2017contexiot} (See Listing~\ref{listing:physicalChannel}). The second example is an app that flashes the lights by adjusting the light intensity and changes the light color at regular intervals. This process creates visual stimuli that can trigger seizures in people who suffer from photosensitive epilepsy~\cite{ronen2016extended}. Similar health-related risks can be inflicted on users through other physical processes such as temperature and sound.
To address misuse of physical processes in these examples, one solution would be to construct a set of templates that define insecure and unsafe device states for side channels and health-related risks. For instance, a template says that an app must not change the volume of a music player above a threshold to prevent hearing loss and tinnitus. The analysis then tracks the device states either at install time or run-time to ensure that an app does not cause the volume state to exceed a threshold or create spikes.

In the third example, an adversary controls a physical process to control some other devices indirectly. For instance, an adversary increases the room's temperature by turning on the heater to activate an app that opens the window when the room temperature exceeds a threshold value~\cite{HuCCS2018, soteria}. This process would allow a burglar to break into homes via windows by controlling the room's temperature. To address indirect access to devices, an app may add extra path conditions to guard device actions based on the app's context. Turning to our example, the window would be open when the temperature value is above a threshold and with some additional conditions such as when the user is at home and when the time is between sunrise and sunset.

\begin{lstlisting}[float=t!, language=Java, caption= An example code block that leaks sensitive information through physical channels, label=listing:physicalChannel, escapeinside={(*}{*)}]
/* An app leaks information by changing light intensity */
/* Similar logic can be used to strobe the light */
subscribe(motion, "motion.inactive", motionInactiveHandler)
def motionInactiveHandler(evt) {
    runIn(60 * minutes, checkMotionStatus)
}
def checkMotionStatus(evt) {
    if (evt.value == "inactive") { // motion inactive
      // setting intensity of the switch 0 
      myLight.setLevel(0)
      changeIntensity()
    }
}
def changeIntensity() {
    def value = myLight.currentState("level")
    // misuse light functionality 
    if (value<=20) {
      state.bool=true
      myLight.setLevel(value+20) }
    if (value>20 && value<80 && state.bool) {
      myLight.setLevel(value+20) }
    if (value>=80) {
      state.bool=false
      myLight.setLevel(value-20) }
    if (value>20 && value<80 && !state.bool) {
      myLight.setLevel(value-20) }
    // change light intensity every 3 seconds 
    runIn(60*0.05,changeIntensity) 
}
\end{lstlisting}

\vspace{3pt}\noindent\textbf{Simulation and Modeling of IoT Programs.} A collection of many IoT devices form a complex system that requires simulators to execute and analyze them accurately. In contrast to traditional modeling and simulation frameworks, simulation of large-scale and heterogeneous IoT environments requires capturing the state of many devices, and the interdependence between events, actions and computational logic~\cite{kecskemeti2017modelling}. The research community and industry have recently explored the requirements for modeling and simulation of IoT implementations~\cite{adjih2015fit, d2016simulation,lee2015modeling, han2014dpwsim}. For instance, IoT-lab provides an infrastructure for testing heterogeneous IoT devices~\cite{adjih2015fit}, and IoTify enables IoT application development by simulating virtual devices in the cloud~\cite{kecskemeti2017modelling}. However, to our knowledge, current IoT simulation tools that researchers often use (\eg SmartThings web-based IoT simulator~\cite{smartthingsSimulator}) have insufficient support for diverse devices and events, which prevents the simulation of apps that have various functionality. 

Another noteworthy point is that physical processes of devices including temperature, illuminance, power consumption, and humidity are often hard to replicate in a simulated environment.  Similar to simulating cyber-physical systems and other physical process-driven systems, IoT analysis tools must consider the evolution of the state of an IoT system over time. This requirement motivates the need for an IoT simulation environment that executes IoT apps by means of a discrete-event simulation engine through continuous-time solvers and state machine-based modeling~\cite{lee2015modeling}.

\vspace{3pt}\noindent\textbf{Automated Test-case Generation.} Dynamic analysis of IoT apps requires input data for execution of the apps~\cite{schwartz2010all}. In IoT apps, inputs are the events that trigger the apps (\ie entry points of an app), and user and device inputs. This introduces a challenge of automating systematic and scalable input generation for IoT apps that control a diverse set of devices with a wide range of internal states. For instance, devices such as a thermostat and power meter may have a discrete (\eg integer-valued) or continuous attributes that would lead to a large input space---generating an input for every possible value in such cases would result in a large number of test cases. 

Similar to other computing platforms, fuzzing and symbolic execution can be used to increase code-coverage for an automated test-case generation. Fuzzing executes the app with random input data, and symbolic execution uses symbolic inputs to perform path-based exploration~\cite{cadar2011symbolic}. For instance, tools for Android, such as Google's Android Monkey~\cite{androidMonkey}, generate random test case inputs of user events and system-level events. To improve test input generation, recent techniques use heuristics that guide input generation to cover app source code intelligently, avoid redundant test paths and enable multi-objective automated testing~\cite{vidas2014a5, rastogi2013appsplayground, carter2016curiousdroid, choudhary2015automated, mao2016sapienz}. Yet, to our knowledge, tools that automate test input data and event generation to execute IoT apps are largely non-existent. This motivates future work to improve test-case generation techniques as applied to IoT.

\begin{figure}[t!]
\begin{center}
\includegraphics[width=0.8\columnwidth]{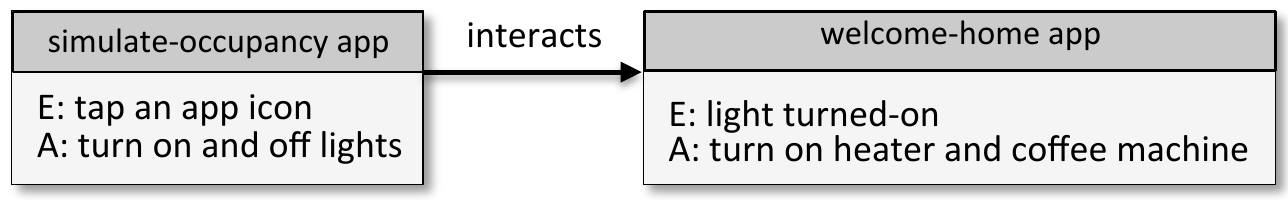}
\caption{An example of interacting IoT apps. \texttt{simulate-occupancy} app interacts with \texttt{welcome-home} app through the light turn-on event. (E is for Event, and A is for Action.)}
\label{fig:interactingApps}
\end{center}
\end{figure}

\vspace{3pt}\noindent\textbf{Multi-app Analysis.} Multi-app analysis that targets IoT environments studies the joint behavior of the apps whereas individual app analysis considers each app in isolation. In a multi-app analysis, apps interact through a common device or abstract events~\cite{soteria, celikIoTGuard}. More specifically, we found that apps interact with each other, (1) when an event handler of an app changes a device attribute,  which triggers another event that is subscribed to by another app; for example, an app turns on the light switch when there is smoke, and another app unlocks the door when the light is turned on, (2) when multiple apps change the same device attribute of some device; for example, a water-leak-detector app shuts off the water valve when there is a leak, while a smoke-alarm app opens the water valve to activate the sprinkler, and (3) when apps that subscribe to the same event change a device attribute in conflicting ways; for example, when motion is detected, one app turns on a switch while another app turns off the switch. We found that apps also interact through \emph{modes}, which are behavior filters that automate device actions. For instance, an app that changes the ``away'' mode to the ``home'' mode when a user arrives home interacts with an app that uses the ``mode change'' event to activate the security alarm. 

The interactions among devices may cause security, safety, and privacy risks even though individual apps are safe in operation~\cite{soteria, HuCCS2018, chi2018cross}. To illustrate, we consider \texttt{simulate-occupancy} app co-resident with \texttt{welcome-home} app (See Figure~\ref{fig:interactingApps}). \texttt{Simulate-occupancy} turns on and turns off the light switch to simulate occupancy when the user is not home. \texttt{Welcome-home} brews coffee and turns on the heater when the light is turned on. \texttt{Simulate-occupancy} interacts with \texttt{welcome-home} through the ``light-on'' event. However, unexpected behavior may happen when these apps interact with each other. In the example, the analysis reveals a safety violation when the user is not home:  the heater and coffee machine are turned on when the bedroom light is turned on because the ``light-on'' is used as an event in \texttt{welcome-home} app. To prevent undesired and unsafe states through interactions, an analysis requires finding the interactions among apps, developing policies for undesired device states, checking the app conforms to those properties when interacting with other apps and blocking the states causing the policy violations.

\begin{figure}[t!]
\begin{center}
\includegraphics[width=0.8\columnwidth]{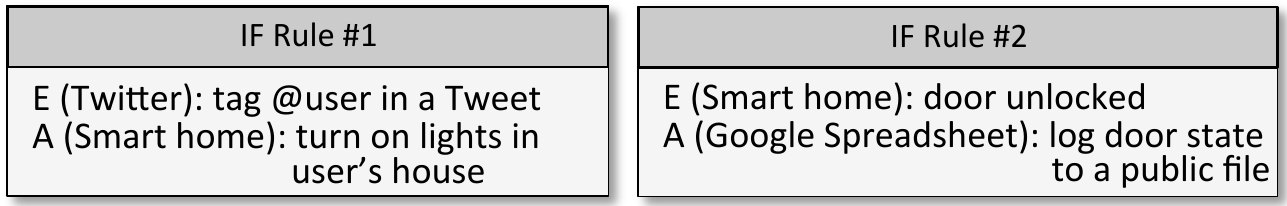}
\caption{Example IF rules in a trigger-action platform. The left IF rule causes an integrity violation, and the right one violates user privacy. (E is for Event, and A is for Action.)}
\label{fig:triggerActionPlatform}
\end{center}
\end{figure}

\vspace{3pt}\noindent\textbf{Interaction between IoT Devices and Trigger-action Platform Services.}  Trigger-action platforms such as IFTTT~\cite{ifttt}, Zapier~\cite{zapier} and Apiant~\cite{apiant} allow users to connect services together. Services include a set of APIs on a trigger-action platform. Users authorize services to their trigger-action platform accounts. For example, a user with a SmartThings IoT platform account can authorize the SmartThings service through the OAuth protocol to communicate with her SmartThings account. Services communicate with each other using REST APIs over HTTP~\cite{fernandes2018decentralized}. Trigger-action platforms allow users to create custom automation on services through DO and IF rules. These rules let users connect a trigger in a service to take the desired action in another service---when an event happens in a service, the platform automatically triggers a separate action in another service. For instance, As of May of 2018, IFTTT has the largest market share~\cite{ifttt-users}; it provides users with 500 services, 158 of which are IoT services. IFTTT enables IoT applications such as fitness trackers and other wearables, hobbyist projects, and connected homes. DO rules act as virtual buttons, which can trigger a set of actions; for example, a DO rule may turn on a smart switch when a button is tapped. IF rules combine two services using a trigger and an action; for example, an IF Rule may make a phone call to the security guard when a motion sensor of a smart home service detects motion after midnight. Users are required to install a companion app provided by the trigger-action platform to trigger DO rules. IF rules run automatically after users configure them via a trigger-action platform web API. 

Similar to multi-app analysis, the interaction between IoT devices and trigger-action platform services may cause security and privacy issues~\cite{celikIoTGuard, surbatovich2017some, musardCCS}. Figure~\ref{fig:triggerActionPlatform} shows two examples of IF rules that connect a smart home to Twitter and Google Spreadsheet services. \texttt{IF rule \#1} turns on lights when a user is tagged in a Twitter post. \texttt{IF rule \#2} logs the door state to a public spreadsheet when the door is unlocked. In the first example, an integrity violation occurs because the untrusted event (Tweet post) changes the state of a trusted action (light on). In the second example, a confidentiality violation occurs because the sensitive information (door unlocked) is made publicly available. Another point worth noting is that these services may also create interactions between IoT apps and services---the actions and events of services and IoT apps can be linked together, similar to the case of interaction between multiple apps. 

Analyses targeting trigger-action platforms require information flow analysis that considers the security and privacy of the environment. More specifically, an analysis may extract the events and actions of the trigger-action rules and label them with the integrity and confidentiality labels. For instance, unlocking a door might be labeled with trusted, and saving a device state to a public file might be labeled confidential. We found that this process is not a trivial endeavor because trigger-action rules are strings and a rule's event and actions often do not match with the capabilities defined in an IoT programming platform. For instance, Santa detector IFTTT rule's definition~\cite{IFTTTsantaDetector} says that ``Ho ho ho! Receive a notification when Santa arrives to deliver you some Merry Christmas joy (and presents)''. Determining the actions and events, and labeling them may need user help or advanced natural language processing techniques.

\subsection{IoT Application Idiosyncrasies}
\label{sec:IoT-app-idiosyncrasies}
Each IoT platform has its own idiosyncrasies based on how they structure the apps and the programming languages they use. These idiosyncrasies require special treatment for analysis precision. In this subsection, we give a couple of example idiosyncrasies.

\begin{lstlisting}[float=t!, language=Java, caption=Sample code blocks for RESTful APIs, label=listing:general-challenges, escapeinside={(*}{*)}]
/* An example use of Restful APIs */
mappings {
  path("/switches") {
    action: [GET: "listSwitches"] }
  path("/switches/:command") {
    action: [PUT: "updateSwitches"] }
}
def listSwitches() {
    switches.each {
      resp << [name: it.displayName, value: 
              it.currentValue("switch")] }
    return resp
}
\end{lstlisting}

\vspace{3pt}\noindent\textbf{RESTful APIs.} RESTful APIs allow external entities to access smart devices and manage those devices. For instance, an app can set the cooling point of a climate control system when the temperature value obtained from a weather forecasting service is above a threshold. These apps declare mappings that relate endpoints, HTTP operations, and callback methods. The SmartThings platform names these apps web-service apps~\cite{SmartThingsWebService}, other platforms provide similar functionality through APIs that enables communicating with the external services. For instance, AWS IoT Core allows both companion and server apps to access connected devices through RESTful APIs~\cite{amazonIoT}. Listing~\ref{listing:general-challenges} shows a code snippet of a SmartThings web-service app. The \texttt{/switches} endpoint handles an HTTP GET request and returns the state information of configured switches by calling the \texttt{listSwitches()} method; the \texttt{/switches/:command} endpoint handles a PUT request by invoking the \texttt{updateSwitches()} method to turn on or off the switches. In our analysis of SmartThings apps, we found 23 official and six third-party web-service apps. These APIs might be used to transmit sensitive data to external services or receive undesired device commands from external services and~\cite{saint-taint-analysis}. Turning to our example app, if an adversary compromises the forecast server and sends fake temperature values to the app, she can turn on many high-power devices to cause outages~\cite{blackIoT}.  

\begin{lstlisting}[float=t!, language=Java, caption=Sample code blocks for language-inherited operations, label=listing:languageBased, escapeinside={(*}{*)}]
/* A code block of an app using closures */
def eventHandler(evt) {
    def currSwitches = switches.currentSwitch  
    def onSwitches = currSwitches.findAll {   
        switchVal -> switchVal == "on" ? true : false
    } 
}
/* Reflection example 1 */
def getMethod() {
  httpGet("http://url") { resp ->
      if (resp.status == 200) {
          name = resp.data.toString()
      }
  }
  "$name"() // call by reflection
}           
def foo() {...}
def bar() {...}
/* Reflection example 2 */
subscribe(presenceSensor, "present", presenceChanged) 
subscribe(presenceSensor, "not present", presenceChanged) 
def presenceChanged(evt) {
    def  s
    if (evt.value == "not present") {
      s = "offDevices"
    } else {
      s = "onDevices"
    }
  performAction(s)
}
def performAction(String f) {
     $f()  // call by reflection 
}
def onDevices() { //  turns on switches }
def offDevices() { // turns off switches }
def otherFunction() { // leak data or misuse device states }
\end{lstlisting}

\vspace{3pt}\noindent\textbf{Language-Inherited Operations.} Analysis techniques need to address the challenges, which programming languages of IoT platforms pose, for analysis precision. In the following discussion, we will provide two examples from IoT apps developed with the Groovy language on the SmartThings platform: closures and call by reflection. We found in our corpus 37 official, and nine third-party SmartThings apps use closures and nine official apps and one third-party app use call by reflection. 
Closures are often used in SmartThings apps to loop through a list of devices and perform computation on each device. Listing~\ref{listing:languageBased} (lines 1--7) shows an example code block in which a closure is used to iterate through the \texttt{currSwitches} object to identify switches that are on. 
For analysis precision, tools need to analyze the structure of closures and inspect expressions within the closures, for example, to see how taints should be propagated in taint tracking~\cite{saint-taint-analysis}.

Call by reflection is used to invoke a method by passing its name as a string. For instance, a method \texttt{foo()} can be invoked by declaring a string \texttt{name="foo"} requested from an external server through the \texttt{httpGet()} interface and thereafter called by reflection through \texttt{\$name} (See Listing~\ref{listing:languageBased} (lines 8--18)). In another example, a developer defines a string conditioned on the state of a presence sensor and passes the string as an argument to a function call (See Listing~\ref{listing:languageBased} (lines 19--36)).
To handle reflective calls, an analysis's call graph construction may add all methods in an app as possible call targets, as a safe over-approximation~\cite{saint-taint-analysis}. For the example in Listing~\ref{listing:languageBased}, an analysis may include both \texttt{foo()} and \texttt{bar()} methods into the targets of the call by reflection in the call graph of an app. Furthermore, an analysis may use string analysis to identify possible values of strings and refine the target sets of reflective calls.

\subsection{Analysis Sensitivities}
\label{sec:analysis-sensitivity}
IoT app analysis can benefit from a more precise program analysis, such as context-sensitive analysis. We next present sensitivities an IoT source code analysis might need for precision and motivate them through code examples. Although these examples are from SmartThings apps, the sensitivity issues are valid for all IoT programming platform apps as many IoT platforms rely on general-purpose programming languages. 

\begin{lstlisting}[float=t!, language=Java, caption=An example code block for  flow sensitivity, label=listing:flow-sensitivity]
energy = powerMeter.currentValue
energy = developer_threshold
message = "energy consumption is $energy"      
sendSMS(message, "attackerPhone")
\end{lstlisting}

\vspace{3pt}\noindent\textbf{Flow Sensitivity.} Flow sensitivity considers the order of execution in a program analysis~\cite{nielson2015principles}. Specifically, a flow-sensitive analysis accounts for variables whose contents change during program execution. In contrast, in a flow-insensitive analysis, a variable includes one qualifier abstracting the values that the variable gets during the entire program execution. In Listing~\ref{listing:flow-sensitivity}, an example IoT app is presented. A flow-sensitive analysis would not flag it to have a data leak, because the message variable has a final value defined by the developer regardless of the sensitive value the power meter has (\texttt{powerMeter.currentValue}). However, a flow-insensitive analysis would flag it to leak sensitive data because it determines that the current value of the power meter can be leaked when the ordering of assignments is not taken into account. 

\begin{lstlisting}[float=t!, language=Java, caption=An example for illustrating context sensitivity, label=listing:context-sensitivity]
def presenceHandler(evt) {
    if (evt.value == "present") {
        take_actions("present")
    } else {
        take_actions("not present")
    }
}

def take_actions(evt_value) { 
    if (evt.value == "present") {
        door.unlock(); lights.on()
        msg = "do not disturb please"
        sendSMS(msg, "userDefinedPhone")
    }
    if (evt.value == "not present") {
        door.lock(); lights.off();
        msg = "user left, event: $evt_value"
        sendSMS(msg, "attackerPhone")
    }
}
\end{lstlisting}

\vspace{3pt}\noindent\textbf{Context Sensitivity.} Context-sensitive analyses span multiple procedures, considering a target function block within the context of the code calling it~\cite{SharirP81Inter}. Specifically, if call-site contexts are used, only execution paths that are feasible by matching calls and returns are considered during analysis. In Listing~\ref{listing:context-sensitivity}, an analysis using depth-one call-site context sensitivity distinguishes the two call sites of \texttt{take\_action} on lines 3 and 5. This means that the analysis analyzes \texttt{take\_action} separately through arguments of ``present'' and ``not\_present'' for those two call sites.  A context-sensitive analysis infers that, for the first call, there is no data leak since \texttt{msg} is sent to a user-defined phone; yet, for the second call, a message is sent to an attacker's phone, which leaks information. In contrast, a context-insensitive analysis considers even infeasible paths in the control flow graph and would decide that both calls leak information. We found that depth-one call-site sensitivity in 230 analyzed apps was precise. Yet, more complex IoT apps might require contexts of greater depth.

\vspace{3pt}\noindent\textbf{Field Sensitivity.} Field-insensitive analysis treats all fields in an object as equivalent~\cite{sridharan2013alias}. IoT apps can use objects for various purposes; for example, SmartThings provides state objects (\texttt{state} and  \texttt{atomicState}) as external storage to persist data across executions. State variables are often used in conditional branches to guard state transitions. In our analysis, we found 74 official and 34 third-party apps declare state variables. Listing~\ref{listing:field-sensitivity} presents an example app using the \texttt{state} object to store a field named \texttt{switchCounter} to track the number of times a switch is turned on. A field-insensitive system would not distinguish \texttt{presenceCounter} from \texttt{switchCounter} (Indeed, the field insensitive analysis would not consider fields at all). A field-sensitive analysis is required to track all fields defined in the \texttt{state} and \texttt{atomicState} objects. For example, the switch-off device state is guarded by the predicate \texttt{state.switchCounter>10}. Furthermore, the analysis may label state variables in predicates as ``state-variables'', indicating they are stored in external data storage.

\begin{lstlisting}[float=t!, language=Java, caption=An example code block for field-sensitivity, label=listing:field-sensitivity]
subscribe(theSwitch, "switch.on", turnedOnHandler)
// initialize switchCounter and presenceCounter to 0
def turnedOnHandler() {
    s_threshold = 10
    state.presenceCounter = state.presenceCounter + 1
    p_counter = state.presenceCounter
    state.switchCounter = state.switchCounter + 1 
    s_counter = state.switchCounter
    if (s_counter > s_threshold) {
        // invoke device actions
    }
    if (p_counter == 1) {
        // send text message
         state.presenceCounter = 0
    }
}
\end{lstlisting}

\begin{lstlisting}[float=t!, language=Java, caption=An example code block for path-sensitivity, label=listing:path-sensitivity]
input "ther", "capability.thermostat"
tempMax = 0                                    
tempMin = 0                                        

if (developerSetPoint < 65) {                         
    tempMin = ther.currentValue            
}
if (developerSetPoint > 65) {                   
    tempMax = tempMin                             
}
message = "thermostat heating is set to: $tempMax"   
sendSMS(message, "attackerPhone")
\end{lstlisting}

\vspace{3pt}\noindent\textbf{Path Sensitivity.} Path sensitivity requires that the predicates at conditional branches are considered in a program analysis~\cite{nielson2015principles}. For instance, in Listing~\ref{listing:path-sensitivity}, the value of the sensitive information \texttt{ther.currentValue} never flows to the \texttt{message} variable because the assignments \texttt{tempMin = ther.currentValue} and \texttt{tempMax = tempMin} never execute together in the program execution.  A path-insensitive system, however,  will conservatively analyze the impossible program execution "\texttt{tempMax = 0}; \texttt{tempMin = 0}; \texttt{tempMin = ther.currentValue}; \texttt{tempMax = tempMin}; \texttt{message = "thermostat... : \$tempMax"}" in which the message string contains sensitive information due to an explicit flow from the thermostat state (\ie \texttt{ther.currentValue}). One way of achieving path sensitivity is through predicate analysis. This is to track the predicates on a particular path during analysis. Take Listing~\ref{listing:dependence-analysis} as an example. There are three feasible paths in \texttt{presentHandler}: (1) \texttt{userTemp=0} and \texttt{currentValue("power")<50} as the path condition of the path that returns constant value 68; (2) \texttt{userTemp=0} and \texttt{currentValue("power")$\geq$50} as the path condition of the path that sends a text message, turns off the switch and returns a constant 63; (3) \texttt{userTemp!=0} as the path condition of the path that returns \texttt{userTemp}.

\begin{lstlisting}[float=t!, language=Java, caption=An example code block for predicate analysis and provenance tracking, label=listing:dependence-analysis]
input "userTemp", "number", title: "Degrees", description: "Adjust temp or default is used by this many degrees", required: false, defaultValue:0
subscribe (presenceSensor, "present", presenceHandler)

def presentHandler() {
    def threshold = 5
    def x =  threshold + evaluate(userTemp)
    thermostat.setHeatingPoint(x)
}

def evaluate() {
    if (userTemp == 0) {
        if (currentValue("power")<50) {
             return 68     
        } else {
             sendSMS(userPhone, "power usage is high")
             lightSwitch.off() // prevent high energy use 
             return 63      
        }
    } else {
        return userTemp     
    }
}
\end{lstlisting}

\vspace{3pt}\noindent\textbf{Provenance tracking.} It is often necessary for an analysis to track sources of data, for example, whether a piece of data is hard-coded by the developer or received as a user input. When such data is used in a device action, knowing its provenance can be extremely helpful in deciding whether the action is intended, or by mistake, or even malicious. In the example of Listing~\ref{listing:dependence-analysis}, constants 63 and 68, and \texttt{threshold} are hard-coded by the developer, and as a result \texttt{x} is computed from hard-coded data by the developer; therefore they should be labeled as ``developer-defined''. In some cases, a user of an application can define some data at install time. For instance, if the threshold value were entered by a user, then \texttt{x} would receive both the label ``user-defined'' and ``developer-defined''. In our analysis of 230 SmartThings apps, we found that apps mostly propagate a developer-defined constant or a user input to places that change device attributes. Occasionally, simple arithmetic is performed; for example, a user input is stored in \texttt{y}, followed by \texttt{x=y+10}, followed by changing a device attribute using \texttt{x}. 

%% file: tableExampleApps.tex
\begin{table}[t!]
\caption{Description of official and third-party SmartThings IoT apps used in our discussion.}
\label{table:sample-apps}
\centering
\def\arraystretch{1.2}
\setlength{\tabcolsep}{10pt}
\resizebox{\textwidth}{!}{%
{\small{
\begin{threeparttable}[b]
\begin{tabular}{l|c|c|c|c|c|c|}
\cline{2-7}
 & \multicolumn{2}{c|}{\textbf{Number of Apps}} & \multicolumn{2}{l|}{\textbf{Unique Device Types}} & \multicolumn{2}{c|}{\textbf{Avg/Max LOC}} \\ \hline
\multicolumn{1}{|l|}{\textbf{App functionality}} & \textbf{Official} & \textbf{Third-party} & \textbf{Official} & \textbf{Third-party} & \textbf{Official} & \textbf{Third-party} \\ \hline \hline
\multicolumn{1}{|l|}{Convenience} & 80 & 26 & \multirow{6}{*}{49} & \multirow{6}{*}{37} & \multirow{6}{*}{244/2633} & \multirow{6}{*}{247/1360} \\ \cline{1-3}
\multicolumn{1}{|l|}{Security and Safety} & 19 & 10 &  &  &  &  \\ \cline{1-3}
\multicolumn{1}{|l|}{Personal Care} & 10 & 0 &  &  &  &  \\ \cline{1-3}
\multicolumn{1}{|l|}{Home Automation} & 48 & 24 &  &  &  &  \\ \cline{1-3}
\multicolumn{1}{|l|}{Entertainment} & 10 & 0 &  &  &  &  \\ \cline{1-3}
\multicolumn{1}{|l|}{Smart Transport} & 1 & 2 &  &  &  &  \\ \hline
\end{tabular}
\begin{tablenotes}
    \item [\textdagger] \small{We determined an app's functionality by checking definition blocks in its source code.}   \hfill \break
\end{tablenotes}
\end{threeparttable}
}}
}
\end{table}

%% file: StudyOfIoTSystems.tex
\section{Study of IoT Analysis Systems}
\label{sec:comparativeAnalysis}
This section presents a study of six recent IoT analysis systems from the literature that use program-analysis techniques for security and privacy. Table~\ref{table:systemSummary} gives an overview of the systems. We begin by introducing analysis techniques used in these systems (Section~\ref{sec:analysisMethods}). The systems, excluding FlowFence, use SmartThings apps for evaluation; thus, we present a background of SmartThings apps (Section~\ref{sec:smartThings-analysis}). Lastly, we study systems with regards to the issues we have introduced (Section~\ref{sec:systemStudy}). In particular, we contrast analysis types and practical implementation specifics. 

\vspace{3pt}\noindent\textbf{IoT Systems.} We give an overview of six recent IoT analysis systems studied throughout. 

\begin{enumerate}
\item\emph{FlowFence} enforces sensitive data flow control in IoT apps and discloses intended data flow patterns to restrict the usage of sensitive data in IoT apps~\cite{fernandes2016flowfence}.

\item\emph{Saint} is a static taint analysis tool that finds sensitive data flows in IoT apps by tracking information flow from taint sources to taint sinks~\cite{saint-taint-analysis}.

\item\emph{ContexIoT} is a context-based permission system that infers the app context automatically and enforces permissions based on that context~\cite{jia2017contexiot}.

\item\emph{SmartAuth} collects device information, annotations and descriptions from app source to generate an authorization interface~\cite{tian2017smartauth}.

\item\emph{ProvThings} captures system-level provenance through security-sensitive APIs and leverages it for forensic reconstruction and attack investigation~\cite{wang2018fear}.

\item\emph{Soteria} extracts a state-model from an IoT app's source code for validating whether an app or multi-app environment adheres to safety, security, and functional properties~\cite{soteria}.
\end{enumerate}

\input{tableIoTSystemsSummary.tex}

\subsection{Fundamental Analysis Techniques}
\label{sec:analysisMethods}
We give an overview of analysis techniques used in six examined IoT analysis systems. Section~\ref{sec:systemStudy} studies the systems with respect to these techniques. 

\vspace{3pt}\noindent\textbf{Taint Tracking.} Taint analysis begins by identifying sensitive data at a taint source with a label that shows the type of information. Taint tracking then starts from a taint source and propagates taint when tainted data is copied and deletes taint when all traces of tainted data are removed (\eg when some variable is loaded with a constant). The impacted data is then flagged at a taint sink (often via the Internet or messaging interface) before it is sent out the system. Lastly, the impacted data is investigated with malware detection tools or by human analysts to determine whether a leak actually constitutes a violation.

\vspace{3pt}\noindent\textbf{Code Instrumentation.} Code instrumentation adds specific code to the source code of an app to collect the app's run-time behavior~\cite{nethercote2004dynamic}. The code added during instrumentation is often called instrumented code. The instrumented code executes as part of the program's normal behavior, but it collects information necessary for some analysis such as context identification, attack detection, and attack reconstruction. Instrumenting every instruction of an app may incur high memory and performance overhead; thus, instrumentation aims to add the minimal code necessary for analysis.

\vspace{3pt}\noindent\textbf{Symbolic Execution.} Symbolic execution indicates that an app is executed with symbolic value as an argument~\cite{baldoni2018survey}.  Unlike concrete execution, where the path is decided by the input, in symbolic execution, the app may practice any feasible path. Symbolic execution enables reasoning about an app behavior on many different inputs, which enables to discover infeasible paths,  identify bugs and vulnerabilities, and create test inputs~\cite{schwartz2010all}.

\vspace{3pt}\noindent\textbf{Model Checking.} Model checking is used to analyze the correctness of software concerning some formally defined program property~\cite{jhala2009software}. Systems or applications are first represented as finite state machines, and the execution of the software is validated against specified specifications through a generic model checker. The specifications are written in temporal logic formulas such as Linear Temporal Logic (LTL) and Computational Tree Logic (CTL)~\cite{clarke1981design}.

\vspace{3pt}\noindent\textbf{Program Slicing.} Program slicing is used to compute program slices that include the program parts affecting the values at some point of interest~\cite{weiser1981program}. For example, the slice of a value at a statement includes a set of statements involved in computing the value in that statement. Program slicing can be used, among others, in debugging to capture the minimal program essentials, and in information flow control to restrict trusted data from interacting with untrusted data. 

\vspace{3pt}\noindent\textbf{Opacified Computing.} Opacified computing provides sandboxes in places where an app have functions that access privacy-sensitive information or device states. Under this model, developers explicitly declare intended functions, and a model is constructed to enforce access to the declared functions and prevent all others~\cite{fernandes2016flowfence}. To achieve this, the developers split an app into modules that operate on functions. The sandbox accumulates information from functions and returns the results which only respect the flow policies.

\subsection{Analysis of SmartThings Apps}  
\label{sec:smartThings-analysis}
The analysis systems, excluding FlowFence, use SmartThings apps for evaluation. We provide a brief overview of SmartThings apps and present techniques for program analysis of its apps.

\vspace{3pt}\noindent\textbf{SmartThings Apps} are developed with a dynamic, object-oriented language Groovy in a sandboxed environment~\cite{fernandes2016security, saint-taint-analysis}. The sandbox limits developers to a specific subset of the Groovy language for performance and security. For instance, the sandbox bans apps from creating their own classes and threads. The cloud backend creates software wrappers for physical devices and runs the apps. SmartThings apps are executed within the SmartThings ecosystem, either in the hub or the SmartThings cloud. Users can install SmartThings apps either from the market or proprietary system via SmartThings Mobile~\cite{soteria}. In the former, publishing an app in the official market requires the developer to submit the source code of the app for review. Official apps appear in the market after the completion of a lengthy review process~\cite{smartThings-review, soteria}. In the latter, organizations can develop an app and make it accessible using the Web IDE. These self-published apps do not receive any official review process and are often shared in the SmartThings official community forum~\cite{Community, soteria}. 

\vspace{3pt}\noindent\textbf{Program Analysis of SmartThings Apps.} Performing a program analysis from the source code of an app requires, among other things, building of the app's Inter-procedural Control Flow Graph (ICFG)~\cite{saint-taint-analysis}. Since Groovy is a JVM-hosted language, one natural approach would be first to compile Groovy code into Java bytecode using the Groovy compiler and then perform analysis via the help of an analysis framework such as Soot~\cite{vallee1999soot}. However, we found that this approach may not be feasible due to the heavy use of reflection in the bytecode generated by the Groovy compiler~\cite{saint-taint-analysis}. In particular, the Groovy compiler translates direct method calls into a call by reflection. IoT systems often analyze Abstract Syntax Tree (AST) representations of Groovy source directly. The Groovy compiler supports customizing the compilation process by supporting compiler hooks, through which one can insert extra passes into the compiler. This is similar to the modular design of the LLVM compiler~\cite{llvm}. Therefore, systems often use \texttt{ASTTransformation} to hook into the compiler, \texttt{GroovyClassVisitor} to obtain the entry points and the structure of the app and \texttt{GroovyCodeVisitor} to visit method calls and expressions inside AST nodes~\cite{GroovyVisitor, saint-taint-analysis}. 

\input{tableSystemsAnalysis.tex}

\subsection{Review of IoT Systems}
\label{sec:systemStudy}
We review the IoT analysis systems in light of the program-analysis issues developed in Section~\ref{sec:prog-analysis-criteria}. We broadly split the systems into two groups based on their goals. The first group includes FlowFence and Saint for privacy, and the second group includes ProvThings, SmartAuth, ContexIoT, and Soteria for safety and security. Our review discusses issues in IoT that have been addressed by prior work and issues that remain open problems. We summarize the characteristics of the systems in Table~\ref{table:comp_analysis}. The following sections discuss the findings of this review process.

\subsubsection{Systems for Privacy} 
We start our analysis with FlowFence and Saint for use (and potential avenues for misuse) of sensitive information in IoT apps. The main difference between FlowFence and Saint lies in the application of the taint tracking. FlowFence, a dynamic system, enforces intended data flow patterns through Quarantined Modules (QMs) whereas Saint, a static system, tracks data flow paths from taint sources to taint sinks. In FlowFence, a developer splits the source code of an app into QMs. QMs run on sensitive data in a sandbox. When a QM accesses sensitive information, taint from data sources (\eg a photo taken by a camera) is tracked, and the data is passed to the sandboxed QM in the form of labeled and immutable data references called opaque handles. Opaque handles can be dereferenced in a QM and transmitted out with a trusted sink API. The data sent through a sink must satisfy a flow policy such as \texttt{<camera, http>} defined in an app's manifest file. In contrast, Saint uses inter- and intra- data flow analysis on IoT apps to find sensitive data flows by tracking information flow from sensitive sources to external sinks. Saint's data flow analysis uses the app's IR. The IR models the app's lifecycle including entry points of an app, devices, user inputs, and call graphs. By leveraging this IR, Saint prunes infeasible paths via path- and context-sensitivity through a work-list based dependency algorithm. 

The other difference between FlowFence and Saint is the implicit flows. The use of QMs in FlowFence eliminates the complexity of handling the implicit flows because non-sensitive code cannot evaluate the value of an opaque handle (return value from a QM) unless it passes to a QM. In contrast, Saint tracks implicit flow by checking the condition of a conditional branch and sees whether it depends on a tainted value. If so, it taints all elements in the conditional branch. 

FlowFence and Saint also differ in addressing IoT-specific issues. Saint addresses SmartThings idiosyncrasies through on-demand algorithms for precision. Yet, for a call by reflection, Saint adds all methods in an app as possible call targets as a safe over-approximation. This increases the number of methods to be analyzed and may lead to over-tainting. FlowFence incurs over-tainting when an app is not accurately separated into QMs. The modulation depends on how a developer structures their data flow controls and IoT-specific mechanisms such as call by reflection. For instance, a developer that does not split an app into the least privilege QMs might cause over-tainting because the analysis does not limit QMs to the code blocks that only process the sensitive data. Another point worth to mention is that FlowFence can track sensitive data flows in multiple IoT apps by enforcing information flow policies between the IoT apps; however, Saint detects sensitive data flows within an individual app.

Lastly, FlowFence's taint tracking requires platform and app developers invest significant efforts towards extending their software to support information flow control, yet Saint automates information flow tracking through backward taint analysis. Both systems require users to make security decisions. FlowFence prompts users for confirmation with taint sources and sinks that indicate how an app will use sensitive data. This may cause frequent flow-prompts to request user permission if publisher policies do not match with the policies. In contrast, Saint presents users with a warning report at install time. The report contains the full data flow paths between taint sources and sinks including the taint labels and taint sink information such as hostname and contact information.

\subsubsection{Systems for Safety and Security}
We study SmartAuth, ContexIoT, ProvThings and Soteria systems designed for safety and security. While these systems differ in analysis precision, runtime and scope, all systems must be responsive to program-analysis issues.

All systems perform analysis on the AST of app source code. In detail, ContexIoT, and ProvThings add instrumentation code to the app source code. Using instrumented code, ContexIoT determines the app functionality under a particular context, ProvThings logs app information for attack investigation and system diagnosis. We note that even though ContexIoT and ProvThings are dynamic systems, they use static analysis to determine where to insert code for obtaining the runtime behavior of apps. Soteria is a static analysis system that extracts a state-model from an app's source code to verify security and safety properties through a model checker. Lastly, SmartAuth performs static analysis to generate an authorization interface for users. SmartAuth complements static analysis with Natural Language Processing (NLP) techniques to capture the differences between an app's actual functionality and the functionality a developer defines. NLP techniques are mainly used to gather data from developer-defined device code annotations and user inputs. For example, the device location is extracted from the device code block, and the app definition is obtained from the definition block of an app's source code. However, the application of NLP techniques might preclude the precise analysis of many practical scenarios. For instance, an app may have incorrect or incomplete device annotations, and some IoT platforms (\eg OpenHAB~\cite{openhab}) do not require an app definition block that can be analyzed.

Systems implement different algorithms for analysis sensitivities depending on their goals. To obtain numerical-valued device attributes through provenance collection, ContexIoT implements taint analysis to find dependencies between numerical attributes. ProvThings computes a backward slice from a numerical-valued attribute as slicing criteria, and Soteria uses dependence analysis to identify a set of possible sources that a numerical-valued attribute can take. To obtain the predicates that guard device actions, ContexIoT gathers the value of the variables on which a device attribute is control-dependent. Soteria uses forward symbolic execution to perform path exploration on source code and accumulates path conditions during exploration. Systems, excluding Soteria, do not track the sources of the values in predicates that show whether a value is defined by a user, hard-coded by the developer, or that user input is modified by the developer. We note that labeling numerical-valued attributes and components in predicates may provide the user with more information for context identification and forensic analysis. For path-sensitivity, Soteria prunes infeasible paths by collecting the predicates at conditional branches and checking whether the conjunction of those predicates is always false. For context-sensitivity, Soteria throws away paths that do not match function calls and returns using depth-one call-site sensitivity. We note that while path- and context-sensitivity is not an issue in ContexIoT and ProvThings, they may add additional instrumented code for provenance and context collection to the infeasible paths.

Systems also differ in handling IoT-specific issues. First, ContexIoT and SmartAuth analyze IoT apps in isolation---collecting context of an individual app;  ProvThings and Soteria, however, capture interactions among apps. ProvThings supports this capability by analyzing provenance graphs of multiple apps, and Soteria constructs a union state-model which represents the unified behavior of apps when they are installed together. Second, systems address SmartThings-specific idiosyncrasies of Restful APIs, closures, and call by reflection in different ways. ContexIoT, ProvThings, and Soteria implement on-demand algorithms for idiosyncrasies; yet systems differ in handling call by reflection. Soteria constructs a call graph by adding all methods as possible call targets of a reflective call and may overapproximate the safety and security violations. ProvThings and ContexIoT instrument all reflective calls, and may perform more instrumentation than needed.

Lastly, some IoT systems require users to make decisions. ContexIoT asks for a user approval of a context through run-time prompts before an action is executed. SmartAuth eliminates this limitation by presenting an authorization interface to users at install time. ProvThings requires users to investigate the provenance graphs and create policies. Soteria defines a set of safety and security properties property through requirements engineering.

%% file: tableIoTSystemsSummary.tex
\begin{table*}[t!]
\caption{A Summary of studied IoT analysis systems.}
\label{table:systemSummary}
\newcolumntype{P}[1]{>{\centering\arraybackslash}p{#1}}
\centering
\renewcommand{\arraystretch}{1.5}
\setlength{\tabcolsep}{2pt}
\resizebox{\textwidth}{!}{
\begin{threeparttable}[b] 
\begin{tabular}{|l|c|c|c|c|c|c|c|c|c|}
\hline
\textbf{System} & \textbf{Purpose} & \textbf{Analysis method} & \textbf{Supp. tech.} & \textbf{Analysis type} & \textbf{IR} & \textbf{Analysis DS} & \textbf{IoT platform} &  \textbf{Input Gen.} & \textbf{\# Apps} \\ \hline \hline
FlowFence~\cite{fernandes2016flowfence} & Data leaks & Opacified comp. & --- & Dynamic & \xmark & Source code & ---\tnote{1} & \cmark\tnote{$\bullet$} & 3 \\ \hline
Saint~\cite{saint-taint-analysis} & Data leaks & Taint analysis & --- & Static & \cmark &  AST & ST\tnote{$\diamond$} & \na\tnote{$\ast$} & 230\tnote{2} \\ \hline
ContexIoT~\cite{jia2017contexiot} & Permission misuse & Code inst. & Taint analysis & Dynamic & \xmark &  AST & ST & \cmark & 283\tnote{3} \\ \hline
SmartAuth~\cite{tian2017smartauth} & Permission misuse & Code inst.\tnote{5} & NLP & Static\tnote{5} & \xmark &  AST & ST & $\ominus$\tnote{$+$} & 180\tnote{4} \\ \hline
ProvThings~\cite{wang2018fear} & Data provenance & Code inst. & Program slicing & Dynamic & \xmark &  AST & ST & \cmark & 236\tnote{3} \\ \hline
Soteria~\cite{soteria} & Abuse prevention &  Symbolic exe.  & Model checking & Static & \cmark &  AST & ST & \na & 65\tnote{2} \\ \hline
\end{tabular}
\begin{tablenotes}[para, footnotesize]
    \item[1] Evaluates three existing IoT apps on Android OS.
    \item[2] Includes both official and third-party apps. \item[3] App type not specified. \item[4] Includes only official apps.
     \item [$\diamond$] ST refers to the SmartThings IoT platform.
     \item [$\ast$] \na, not applicable for a static system.\\
    \item[5] SmartAuth extracts an app's behavior through static analysis; however it also collects runtime information to block unauthorized device actions.
    \\
    \item[$\bullet$] FlowFence, ContexIoT and ProvThings employ brute-force fuzzing that randomly generates user inputs and  events to execute the apps.\\
    \item [$+$] $\ominus$ means that we could not find enough implementation details to be conclusive.
\end{tablenotes}
\end{threeparttable}
}
\end{table*}

%% file: tableSystemsAnalysis.tex
\begin{table*}[t!]
\caption{Review of IoT analysis systems based on our discussion in Section~\ref{sec:prog-analysis-criteria}.}
\label{table:comp_analysis}
\newcolumntype{P}[1]{>{\centering\arraybackslash}p{#1}}
\centering
\renewcommand{\arraystretch}{1.4}
\resizebox{\textwidth}{!}{
\begin{threeparttable}[b]  
\begin{tabular}{P{2cm} | P{1.35cm}  | P{1.35cm} | P{1.35cm}  | P{1.35cm}  | P{1.35cm}  | P{1.35cm}  | P{1.35cm}  | P{1.35cm}  | P{1.35cm}  | P{1.35cm} | P{1.35cm} | P{1.35cm} |}
\cline{2-13}
& \multicolumn{4}{c|}{\textbf{IoT-specific issues}}             
& \multicolumn{3}{c|}{\textbf{IoT app idiosyncrasies}}           
& \multicolumn{5}{c|}{\textbf{Analysis sensitivity}} \\ \hline\hline
\multicolumn{1}{|c|}{\textbf{System}}& \textbf{I.1} & \textbf{I.2}& \textbf{I.3} & \textbf{I.4} 
& \textbf{S.1} & \textbf{S.2} & \textbf{S.3} 
& \textbf{P.1} & \textbf{P.2} & \textbf{P.3} & \textbf{P.4} & \textbf{P.5}  \\ \hline
\multicolumn{1}{|l|}{FlowFence~\cite{fernandes2016flowfence}} 
& \cmark &  \xmark & \cmark & \cmark 
& \na\tnote{\textdagger}  & \na  & \na      
& \cmark & \na & \na & \na & \na  \\ \hline
\multicolumn{1}{|l|}{Saint~\cite{saint-taint-analysis}}       
& \xmark & \xmark & \cmark & \cmark 
& \cmark & \cmark & \cmark 
& \cmark & \cmark & \cmark & \cmark & \cmark  \\ \hline
\multicolumn{1}{|l|}{ContexIoT~\cite{jia2017contexiot}}       
& \xmark & \xmark & \xmark & \xmark 
& \cmark & \cmark & \cmark 
& \cmark & \na & \cmark & \na & \cmark  \\ \hline
\multicolumn{1}{|l|}{SmartAuth~\cite{tian2017smartauth}}     
& \xmark & \xmark & \cmark & \cmark 
& \cmark & \xmark & \xmark 
& \cmark & \xmark & \xmark & \xmark & \xmark  \\ \hline 
\multicolumn{1}{|l|}{ProvThings~\cite{wang2018fear}}       
& \cmark & \xmark & \xmark & \cmark 
& \cmark & \cmark & \cmark 
& \cmark & \na & \cmark & \na & \cmark  \\ \hline 
\multicolumn{1}{|l|}{Soteria~\cite{soteria}}                 
& \cmark & \xmark & \cmark & \cmark 
& \xmark & \cmark & \cmark 
& \cmark & \cmark & \cmark & \cmark & \cmark  \\ \hline
\end{tabular}
\setlength{\tabcolsep}{9.9pt}
\renewcommand{\arraystretch}{1.35}
\begin{tabular}{lll|l|l}
\hline
\multicolumn{5}{|c|}{\textbf{Legend}} \\ \hline 
\multicolumn{2}{|c|}{\textbf{IoT-specific issues}} & \multicolumn{1}{c|}{\textbf{IoT app idiosyncrasies}\tnote{\textdaggerdbl}} & \multicolumn{2}{c|}{\textbf{Analysis sensitivity}} \\ \hline\hline
\multicolumn{1}{|l}{I.1 Multi-app analysis} 
& \multicolumn{1}{|l}{I.2 Trigger-action platform support} 
& \multicolumn{1}{|l}{S.1 RESTful APIs\tnote{$\star$}} 
& \multicolumn{1}{|l}{P.1 Flow sensitivity} 
& \multicolumn{1}{|l|}{P.2 Context sensitivity} \\ \hline
\multicolumn{1}{|l}{I.3 Proactive defense} 
& \multicolumn{1}{|l}{I.4 No runtime prompts}
& \multicolumn{1}{|l}{S.2 Closures and other operations} 
& \multicolumn{1}{|l}{P.3 Field sensitivity}
& \multicolumn{1}{|l|}{P.4 Path Sensitivity}\\ \hline
& \multicolumn{1}{l}{}
& \multicolumn{1}{|l}{S.3 Calls by reflection}
& \multicolumn{1}{|l|}{P.5 Provenance Tracking}
\\\cline{3-4}
\end{tabular}
\begin{tablenotes}[para, footnotesize]
    \item [\textdaggerdbl] We split the criterion of language-inherited operations of the SmartThings platform into ``closures and other operations'', and ``call by reflection''.\newline
    \item [$\ast$] \na~ means that not applicable.
    \item [$\star$] RESTful APIs refer to web-service apps in SmartThings platform. 
\end{tablenotes} 
\end{threeparttable}
}
\end{table*}

%% file: conclusions.tex
\section{Takeaways and Conclusions}
The security and privacy of IoT is a new and emergent area. This work attempts to study IoT application security and privacy research through program-analysis techniques. We began by surveying five major IoT programming platforms to gain insights into the structure of their apps and map their app structures into common building blocks. By studying these IoT platforms, we have distilled the key aspects of program analysis under IoT-specific analysis issues, IoT app idiosyncrasies, and analysis sensitivities. Lastly, we have explored IoT app analysis academic papers over the past two years that employ program-analysis techniques for security and privacy issues. Broadly speaking, most attempts to date focus on issues such as sensitive data leaks, abuse prevention, permission misuse, and provenance collection. Our study yields a natural structure for reasoning about the capacity of the IoT systems and reveals the extent to which each system identifies and mitigates safety, security and privacy issues.

Our key findings through these explorations include: (1) The dominant IoT programming platforms structure their apps around a sensor-computation-actuator idiom, (2) a suite of analysis tools and algorithms targeted at diverse IoT platforms is at this time largely absent, (3) because IoT applications control physical processes through devices, the security and privacy issues are more subtle and difficult to identify, (4) most approaches lack multiple analysis sensitivities such as path- and context-sensitivity, (5) most approaches often do not consider security and safety problems in multi-app environments and through information flows in trigger-action platforms, (6) members of the research community often use the SmartThings platform to evaluate their tools as numerous open-source official and third-party apps are available, and (7) IoT systems often implement algorithms on the Abstract Syntax Tree (AST) of a SmartThings app because of the constraints on Groovy language and proprietary back-end libraries.

While the research community has been effective in providing tools that identify security and privacy issues in specific IoT implementations, many areas remain open problems, and IoT program analysis needs additional progress before apps are safe for broader use.  
First, IoT analysis systems that use program analysis techniques for security and privacy often focus on smart homes. Yet, IoT applications are diverse in terms of type and the number of connected devices. Therefore, the analysis must be responsive to the unique characteristics and constraints of different IoT domains.
Second, current IoT analysis systems may possess scalability problem oft-encountered in formal program analysis in systems such as smart automobiles and industrial IoT, which have a large amount of code for analysis. The research community must consider the practicality of their approaches in IoT systems where large-scale programs are developed and updated on a regular basis.
Third, physical processes in IoT can have effects on critical infrastructure. For instance, IoT devices are able to change power demand in critical infrastructure, accelerate a motor to a velocity, and decrease water usage in an industrial system. This inter-tangled environment expands security issues through subtle interactions between IoT systems and other environments. Therefore, the interactions between systems must be carefully studied to uncover potential security issues.
Fourth, analysis systems often do not assess the impact of approaches on the system resources. Thus, existing IoT solutions may incur high computational cost and energy consumption which might be infeasible for real systems. For instance, an IoT analysis system may need to poll sensor data periodically to obtain device states. The polling could consume sensor battery when the intervals are too short or may limit real-time detection when the intervals are too long. Program analysis and statistical modeling techniques can be combined to create efficient methods to reduce the energy consumption of devices.
Lastly, approaches need to consider taking the right course of action when a security and safety violation happens. Simply blocking a device state or asking a user for approval through runtime prompts could be dangerous. For example, door-unlock action in an app that unlocks the door when there is smoke in the house may not be permitted by the policy or may be asked a user to approve the action. However, dropping the action or no response from a user will result in a locked door, which is potentially unsafe depending on the circumstances. To help keep the IoT environment stable when a violation is detected, several response disciplines can be implemented to preserve the integrity of the environment. 

We envision these explorations to be a central pillar for applying program-analysis techniques to IoT, and providing researchers with insights useful for future work.